\documentclass[11pt,fleqn]{article}

\usepackage[margin=1in]{geometry}
\usepackage{wrapfig}

\newif\ifpdf
\ifx\pdfoutput\undefined
\pdffalse 
\else
\pdfoutput=1 
\pdftrue
\fi

\ifpdf
\usepackage[pdftex]{graphicx}
\else
\usepackage{graphicx}
\fi

\usepackage[english]{babel}

\newcommand{\br}{{\bf{r}}}
\newcommand{\bR}{{\bf{R}}}

\newcommand{\bx}{{\bf{x}}}
\newcommand{\bX}{{\bf X}}
\newcommand{\bY}{{\bf Y}}

\newcommand{{\cis}}{{\em cis}}

\newcommand{\kBT}{{k_{\rm B} T}}

\newcommand{{\trans}}{{\em trans}}

\begin{document}

\baselineskip=14pt plus 3pt minus 2pt
\lineskip=3pt minus 1 pt
\lineskiplimit=2pt
\parskip=8pt plus 1pt minus .5pt


\begin{center}
\topmargin=2in

{\Large\bf Ion Binding Sites and their Representations by 

Quasichemical Reduced Models}

\vspace{0.2in}

{\sf Beno\^it Roux } \\

{\it Department of Biochemistry and Molecular Biology \\ 
           Gordon Center for Integrative Science \\ 
           University of Chicago \\ 
           Chicago, IL 60637 \\ 
           Phone: (773) 493-5303 \\ 
           Email: roux{@}uchicago.edu }


\end{center}

\subsection*{Abstract}

The binding of small metal ions to complex macromolecular structures is typically 
dominated by strong local interactions of the ion with its nearest ligands.
For this reason, it is often possible to understand the microscopic origin of ion binding selectivity by 
considering simplified reduced models comprised of only the nearest ion-coordinating ligands.
Although the main ingredients underlying simplified reduced models are intuitively clear, a formal statistical mechanical 
treatment is nonetheless necessary in order to draw meaningful conclusions about complex macromolecular systems.
By construction, reduced models only treat the ion and the nearest coordinating ligands explicitly. 
The influence of the missing atoms from the protein or the solvent is incorporated indirectly.
Quasi-chemical theory offers one example of how to carry out such a separation in the 
case of ion solvation in bulk liquids, and in several ways, a statistical mechanical formulation of 
reduced binding site models for macromolecules is expected to follow a similar route.
Here, some critical issues with recent theories of reduced binding site models are examined.


\section*{1. Introduction}

Small metal ions are a fundamental component to the structure and function of biological systems.  
Although detailed computation have a central role to play in trying to understand the molecular determinants of ion selectivity
in these complex systems, progress may be achieved by pursuing 
theoretical studies based on simplified reduced models comprised of only the nearest ion-coordinating ligands
\cite{Noskov-2004, Asthagiri-2006, Noskov-2007, Varma-2007, Varma-2008b, Noskov-2008, Dixit-2009, Bostick-2009, Roux-2010, Yu-PNAS-2010, Yu-2011, Rogers-2011, Roux-2011}.
By construction, such reduced binding site models only treat the ion and the nearest coordinating ligands explicitly,
while the influence of the missing atoms from the surroundings (protein or solvent) is incorporated indirectly.
Although the general idea is conceptually simple, a rigorous statistical mechanical treatment is necessary 
to draw meaningful conclusions about complex macromolecular systems.
A typical starting point is to separate the system into  ``inner'' and ``outer'' regions. 
Quasi-chemical theory (QCT) offers one example of how to carry out this type of formal 
separation in the case of an ion in bulk solvent \cite{QCT-Pratt-1998,Pratt-book,Roux-QCT-2010}.
The system is rigorously partitioned into a small fixed spherical volume comprising the ion and $n$ solvent molecules, 
and a second region corresponding to the remaining bulk liquid phase \cite{QCT-Pratt-1998,Pratt-book,Roux-QCT-2010}.
Thermodynamic ion solvation properties are then expressed as a sum over a series of 
clusters comprising the ion and $n$ water molecules, each treated at the QM level, and each explicitly weighted by the probability $P(n)$. 
QCT was shown to be formally equivalent to a treatment of ion solvation based on the small 
system grand canonical ensemble (SSGCE) \cite{Roux-QCT-2010}.
A particularly simple form of QCT, referred to as the primitive QCT (pQCT),
evaluates configurational integrals using the rigid rotor harmonic oscillator (RRHO) approximation.
The studies of Varma and Rempe have relied on pQCT \cite{Varma-2007,Varma-2008b}, 
while those of Bostick and Brooks were formulated from SSGCE \cite{Bostick-2009}.
Both approaches are constructed by first considering the probability of finding $n$ ligands in a spherical region,
a standard procedure in the development of theories of solvation in bulk liquids \cite{Reiss-1998, Pratt-book, Roux-QCT-2010}.
However, both approaches encounter similar difficulties because the statistical properties of the ligands of a protein binding site 
were assimilated with those of a bulk liquid.  
In the following, we critically examine some of the issues arising in the context of QCT and SSGCE.

\section*{2. Analysis }

\subsection*{a) Primitive Quasi-Chemical Theory (pQCT) }

The primitive form of QCT (pQCT) was used to examine the factors governing selectivity in ion channels \cite{Varma-2007,Varma-2008b}.  
To clarify the significance of pQCT when applied to ion channels, it is useful to recall the underlying theoretical framework. 
QCT is a statistical mechanical theory in which the influence of the solvent molecules surrounding an 
ion in a bulk liquid is separated into near and distant contributions \cite{Pratt-book, Roux-QCT-2010}.
The construction of pQCT of ion solvation is illustrated schematically in Fig. 1A.
\begin{figure}[h]
\begin{center}
\includegraphics[width=6.5in]{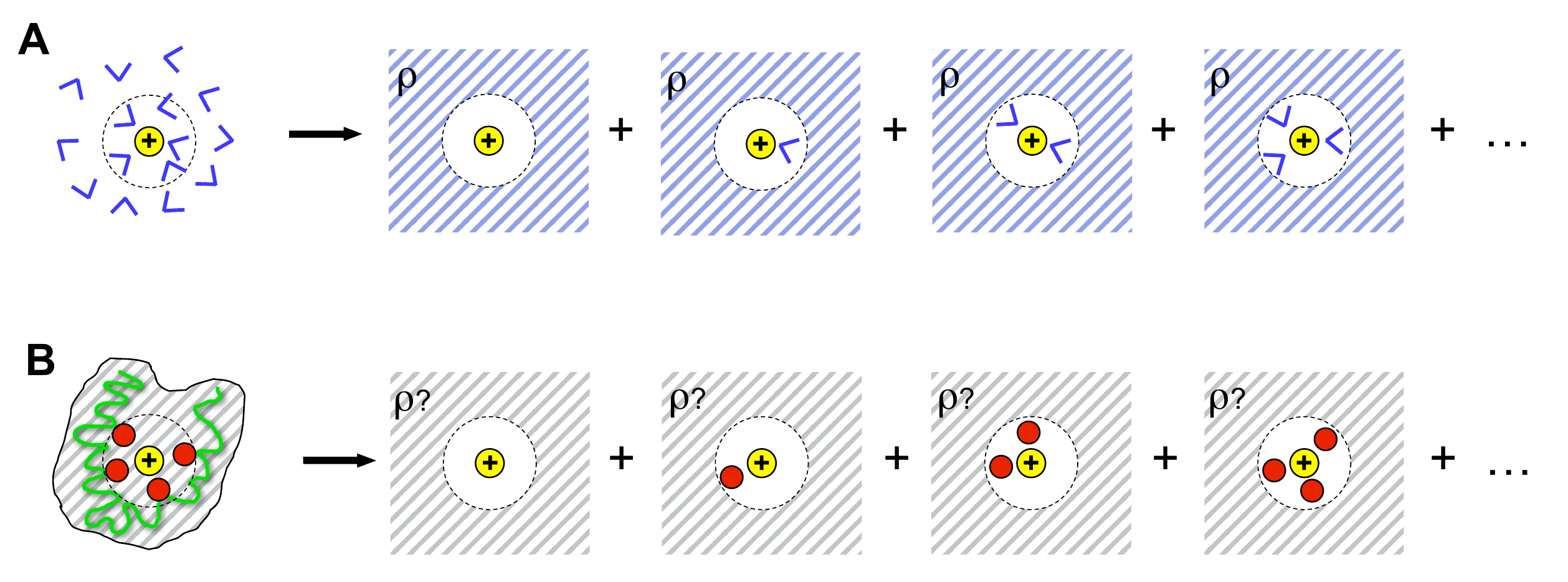}
\end{center}
\caption{   
\footnotesize
Schematic illustrations of primitive quasi-chemical theory (pQCT) in different situations.
(A) Application of pQCT to the hydration of a small ion.  
One the left is shown an ion in bulk water. The dashed line indicates the position of the inner-shell region.
On the right, the pQCT cluster sum appearing in Eq.~(\ref{eq:pQCT}) is depicted for $n=0,1, 2, 3, \dots$.
The water molecules surrounding the ion in the inner-shell are represented explicitly, whereas 
those in the outer shell are represented as a continuum of density $\rho$.
(B) Application of pQCT to an ion binding site in a protein
On the left is shown the ion with its coordinating ligands in the protein binding site.
The ligands are attached to the protein, which is a covalently connected structure,
On the right, a putative pQCT construct of the same system is depicted, 
with the ion-coordinating ligands explicitly represented and 
the outer region incorporated implicitly as a generic liquid continuum (of density ``$\rho$?'').
}
\end{figure}

Central to QCT is the definition of a spherical inner-shell region of radius $R$ centered on the ion. 
A possible route for deriving QCT is to sort out the multitude of possible 
configurational integrals according to the number of solvent molecules occupying the inner-shell region, 
and then treat the latter as a small open system in equilibrium with a bulk region (density $\rho$) 
composed of identical solvent molecules that are free to exchange position with the bulk \cite{Roux-QCT-2010}. 
The statistical mechanical developments bear many similarities to the concept of SSGCE considered by Reiss and co-workers \cite{Reiss-1998}. 
QCT adopts a particularly simple form once it is assumed that the clusters undergo small harmonic fluctuations around a single energy minimum. 
This leads to the so-called ``primitive'' QCT approximation (pQCT), which provides the following expression for the 
solvation free energy of an ion $i$ in a liquid \cite{Roux-QCT-2010}, 
\begin{equation}
G_i^{\rm bulk}  = -\kBT \ln \left [ P_i^0(R_i) \, \sum_n   \, (\rho)^n   \, e^{-\beta \left [E_i^{\rm min}(n) + \Delta W_i - n \Delta \mu \right ] } \, 
                                K_i^{\rm trv}(n)  \right ] 
\label{eq:pQCT}
\end{equation}
where the sum pertains to small clusters comprising one ion surrounded by $n$ solvent molecules. 
Here, $P_i^0(R_i)$ is the probability of finding an empty spherical cavity of radius $R_i$ in the bulk liquid, 
$\rho$ is the density of the bulk liquid, 
$\Delta \mu$ is the excess chemical potential of a solvent molecules in the bulk liquid,
$E_i^{\rm min}(n)$ is the binding energy minimum of the cluster (optimal energy-minimized geometry), 
$\Delta W_i$ is the solvation free energy of the cluster (coupling with the surrounding bulk phase),
and $K_i^{\rm trv}(n)$ is a product of standard translation-rotation-vibration partition functions for the cluster in the gas phase.
Quantities such as $\Delta \mu$ and $\Delta W_i$ are often approximated via a continuum dielectric model
\cite{Rempe-2000,Rempe-2001,Rempe-2004, Roux-QCT-2010}, although this is not necessary.

pQCT was utilized in several studies of hydration of Li$^+$, Na$^+$ and K$^+$ \cite{Rempe-2000,Rempe-2001,Rempe-2004},
for which it was specifically designed.  
Those studies, which combined pQCT with quantum mechanical methods, are of great interest. 
Nevertheless, they did not provide a quantitative assessment of the accuracy of pQCT.
To achieve this goal, we compared the outcome of pQCT with ``exact'' results from MD simulations using 
simple classical models of water and ions \cite{Roux-QCT-2010}.
The comparison shows that the results of pQCT about the hydration of those small metal cations must be interpreted with caution
because the accuracy of the approximation varies for ions of different size \cite{Roux-QCT-2010}; 
Li$^+$ is represented with excellent quantitative accuracy, 
while Na$^+$ is described less accurately and K$^+$ encounters some serious difficulties. 
These tests showed that even when the pQCT approximation yields reasonable solvation free energies, 
it could be quantitatively unreliable regarding average coordination numbers \cite{Roux-QCT-2010}.

Although pQCT and Eq.~(\ref{eq:pQCT}) were derived to describe ion solvation in a bulk liquid,
the approach was also used to study ion selectivity in proteins binding sites \cite{Varma-2007,Varma-2008b}.  
As illustrated schematically in Fig. 1B, 
the application of pQCT to proteins raises a number of issues because this is a situation 
for which many of the assumptions used to derive Eq.~(\ref{eq:pQCT}) are not satisfied.
For example, the mathematical developments leading to Eq.~(\ref{eq:pQCT}) specifically rely on the  fact that solvent molecules 
are {\em identical} and {\em free to exchange} with one another in the bulk phase (see Eqs. 6-13 in ref \cite{Roux-QCT-2010}).
Eq.~(\ref{eq:pQCT}) is an expression that is relevant to an ion in bulk solvent, not in a protein binding site.
The only information about the protein structure that can be incorporated into pQCT is to restrict the 
sum over $n$ in Eq.~(\ref{eq:pQCT}) to a single coordination number.
An application of pQCT and Eq.~(\ref{eq:pQCT}) to discuss ion binding to a protein 
such as the selectivity filter of the KcsA K$^+$ channel is thus of unclear significance 
because the coordinating ligands are tethered to the polypeptide backbone. 
In practice, to apply Eq.~(\ref{eq:pQCT}) to KcsA, one must ``map'' the properties of the selectivity filter onto those of a bulk liquid,
i.e., $\rho$ and $\Delta \mu$, the density and excess chemical potential.
In their study of the KcsA channel, acetamide molecules were used to represent 
the backbone carbonyl groups of the selectivity filter \cite{Varma-2007}, a chemically reasonable choice.
But what value of $\rho$ and $\Delta \mu$, should be used in Eq.~(\ref{eq:pQCT}) 
to best represent the selectivity filter of the KcsA channel is unclear. 
Should one estimate those parameters from the properties of liquid acetamide? 
Or should one try to derive some effective value of those parameters mimicking the interior of an average membrane protein? 
How is the influence of the relative flexibility of the surrounding protein structure incorporated into the theory?
Furthermore, the results depend on the radius of the inner shell, $R_i$, chosen for a particular ion.
These critical issues were not addressed in previous studies of ion selectivity based on pQCT \cite{Varma-2007}.

\subsection*{b) Attempts to extend QCT to inhomogeneous systems}

\begin{figure}[h!]
\begin{center}
\vspace{0.1in}
\includegraphics[width=5.0in]{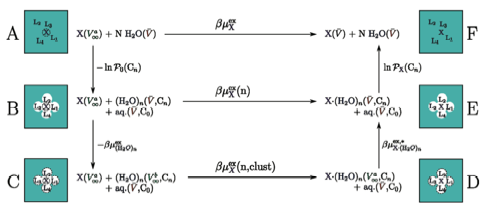}
\end{center}
\caption{   
\label{fig:rempe}
\footnotesize
Schematic illustration of an extension of quasi-chemical theory (QCT) 
for inhomogeneous systems applicable to protein binding sites 
(directly taken from the Figure 2 of reference \cite{Rogers-2011}).
The intermediate stages shown schematically are specified 
by enclosing noninteracting molecules with dotted lines, 
and indicating coordination and excluded volume conditions by shading.  
While this was not specified, it is presumed that the green background represents the 
inhomogeneous environment (including that of the protein) surrounding the ligands and the ion.
}
\end{figure}

An extension of QCT for inhomogeneous systems applicable to protein binding sites was recently proposed \cite{Rogers-2011}. 
The framework was only elaborated schematically in the form of the thermodynamic cycle shown in Fig.~\ref{fig:rempe}.
To facilitate a discussion of ion binding thermodynamics, it is therefore necessary 
to translate the states appearing in this scheme into standard mathematical statistical mechanical expressions
(all configurational integrals will be written in the NVT ensemble for the sake of simplicity). 

The main difference from previous versions of QCT developed for homogeneous fluids is the change 
in the ordering of ligand removal and prearrangement of the coordination structure constraint function. 
The non-covalent transfer of an ion into the protein binding site corresponds the equilibrium steps A$\rightarrow$F.
At its base, the scheme is design to contain a gas-phase ion-ligands cluster formation free energy 
in step step C$\rightarrow$D that is typical of QCT.
The schematic thermodynamic cycles is designed to permit use of gas-phase QM calculations for cluster formation energies.
In the cycle, the $n$-coordination structure specified by $C_n$ is formed, then 
its interaction with the surrounding medium are removed to bring 
it into the gas phase before turning on solute-to-ligand interactions. 
The ambiguity of ``the complex is formed'' must be resolved in 
practice with an indicator function, $I(\bR;Cn)$, which is 1 whenever 
a particular point in phase space satisfies specific local structural requirements and zero otherwise. 
The indicator function is designed to include constraints of coordination and excluded volume conditions 
(i.e., preventing moieties from the surroundings to directly interfere with the specified $n$-coordination structure).  
In Fig.~\ref{fig:rempe}, this is indicated by ``white shading'', but is left otherwise unspecified.
There is a wide range of possibilities to construct practical indicator function.
For example, coordination might be defined by counting the ligands within a maximum distance of 3.5 {\AA} away from the ion. 
Alternately, one may choose a definition  that is based on a maximum root mean squared 
deviation tolerance (RMSD) with respect to an ideal coordination complex geometry.
Different constraints can be defined for each coordination complex of interest for a given system. 
This generalization is appropriate for discussing chemical reactions occurring in arbitrary, inhomogeneous environments 
due to the natural appearance of the environmental potential of mean force (PMF) 
when calculating partition functions with coordination constraints. 

\begin{wrapfigure}[9]{r}{8cm}
\begin{center}
\vspace{-1.1cm}
\includegraphics[width=8cm]{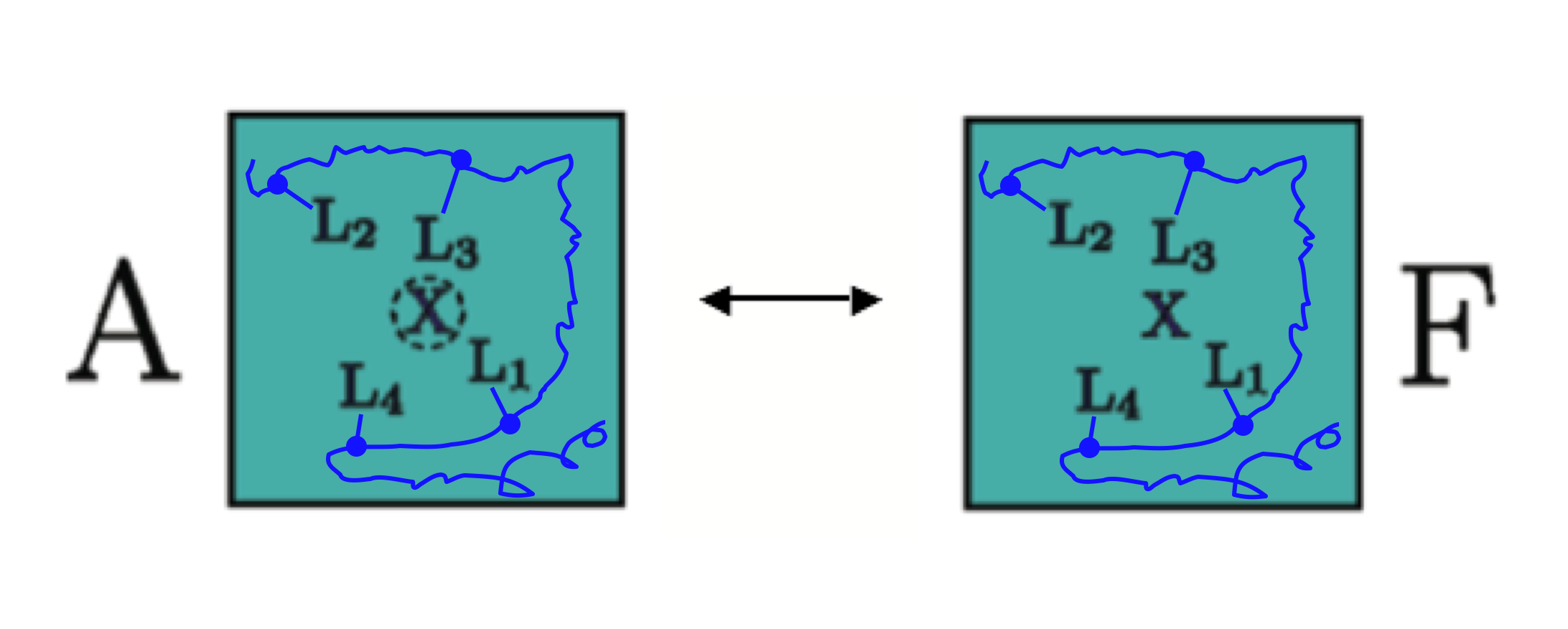}
\label{fig:2b}
\vspace{-0.4cm}
\caption{   
\label{fig:af}
\footnotesize
Equilibrium binding of one ion into a binding site with the ligands L$_{\alpha}$.
The dotted lines around the ion X in state A indicates that the particle is noninteracting.
The blue line represents the protein structure to which the ligands are covalently bonded.  
Adapted from the Figure 2 of reference \cite{Rogers-2011}.
}
\end{center}
\end{wrapfigure}
It is helpful to first consider the equilibrium between the end-states A and F depicted in 
Fig.~\ref{fig:rempe} summarized in Fig.~\ref{fig:af}.
Here, a ``blue line'' was added to remind ourselves that a macromolecule is part 
of the inhomogeneous ``green background'' environment, and that the ligands are covalently tethered to it.
The state A is meant to represent  the configurational integral $Z_A$ 
in which the ion is noninteracting ion,
\begin{equation}
Z_A  =   \int_{\rm apo} d\bR \, e^{-\beta U_0(\bR)}
\end{equation}
where $\bR$ represents all the coordinates in the system, 
and $U_0$ is the potential energy when the ion is noninteracting. 
The state F is meant to represent the configurational integral $Z_F$ of the ion in the protein binding site,
\begin{equation}
Z_F  =  \int_{\rm holo} d\bR \, e^{-\beta U(\bR)} 
\end{equation}
The actual equilibrium binding constant of the ion, $K_{\rm b}$, can be expressed in terms of the ratio $Z_F/Z_A$ as,
\begin{eqnarray}
K_{\rm b} = V { Z_F \over Z_A } \; e^{\beta G_{\rm bulk}} 
\end{eqnarray}
where $G_{\rm bulk}$ is the free energy to transfer the ion X from 
the bulk solution to the gas phase (absolute hydration free energy of the ion), and $V$ is 
the volume of the system spanned by the noninteracting ion ($V = \int d\br_{\rm ion}$).

Four additional intermediate states, B-E, are inserted between the end-point states A and F in Fig.~\ref{fig:rempe}.
They correspond to arbitrary constructs used to define a step-by-step procedure to 
go between the two end-states A and F, and thus compute the total binding free energy of the ion.
To facilitate the analysis, we distinguish three main atomic and molecular components 
in the total potential energy of the system $U(\bR)$: 
the ion (i), the ion-coordinating ligands (l), and the rest of the system (r).
The latter includes the rest of the protein as well as the surrounding bulk solvent.
The system is separated into an inner and an outer region.
The ion and the $n$ coordinating ligands are part of the inner region, while the rest of the system is part of the outer region.
Let the coordinates of the ion and the $n$ coordinating ligand be represented by 
$\bX \equiv \{\br_{\rm ion}, \bx_1,\dots,\bx_n \}$, 
and those for the remaining atoms be represented by $\bY$, and $\bR \equiv \bX, \bY$.
The total potential energy $U$ follows naturally from this decomposition and can be expressed as the sum:
\begin{eqnarray}
U(\bR) &=& u_{\rm il}(\bX) + u_{\rm ir}(\bX,\bY) + \sum_{i=1}^n u^i_{\rm l}(\bx_i) +
            u_{\rm ll}(\bX) +    u_{\rm lr}(\bX,\bY) + u_{\rm rr}(\bY)
\label{eq:sum}
\end{eqnarray}
where the separate energy terms correspond to: 
ion-ligand (il), ion-rest (ir), 
isolated ligand (l), ligand-ligand (ll), 
ligand-rest (lr), and rest-rest (rr).
Such a form arises naturally for a pairwise decomposable molecular mechanical force field, 
although it is always possible to formally re-construct the total potential energy as such a sum of separate terms, 
even in the context of a non-pairwise additive quantum mechanical energy surface. 
We can now try to express mathematically all the intermediate steps, B, C, D, E of Fig.~\ref{fig:rempe}
introduced in ref \cite{Rogers-2011}.

\begin{wrapfigure}[4]{r}{4cm}
\begin{center}
\vspace{-1.1cm}
\includegraphics[width=4cm]{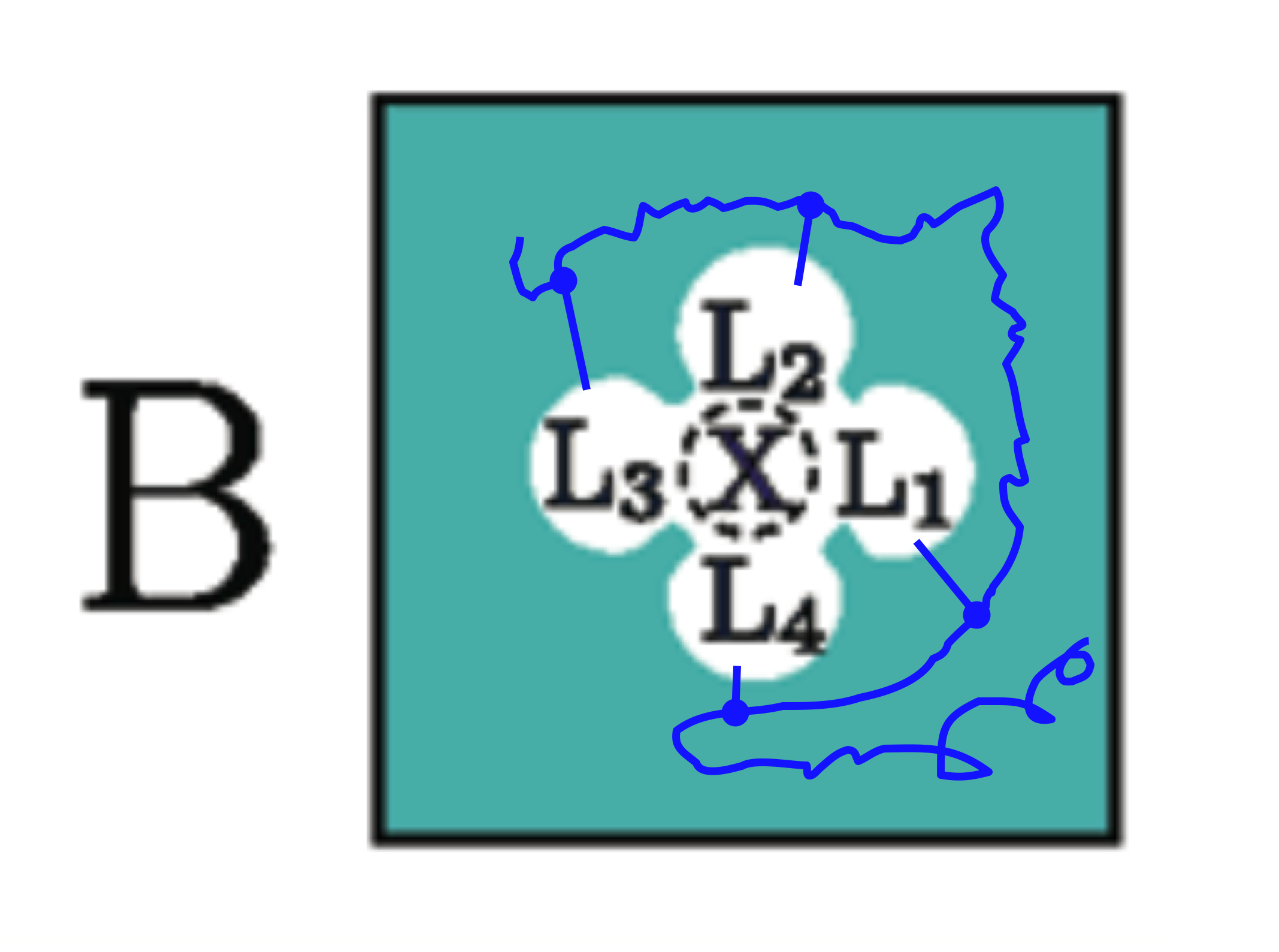}
\label{fig:2b}
\vspace{-0.4cm}
\end{center}
\end{wrapfigure}
State B has both the ion X and the ligands being constrained within coordination 
and excluded volume conditions (surrounded by white shaded area).
The ion X (surrounded by dotted line) is noninteracting and decoupled from its surroundings, while the ligands are interacting normally.
The configurational integral $Z_B$ is,
\begin{equation}
Z_B =  \int d\bX \; d\bY \, I(\bR; C_n) ~ e^{- \beta \left [ \sum_{i=1}^n u^i_{\rm l}(\bx_i) + u_{\rm ll}(\bX) + u_{\rm lr}(\bX,\bY) +u_{\rm rr}(\bY)\right ] }  
\end{equation}

\begin{wrapfigure}[4]{r}{4cm}
\begin{center}
\vspace{-1.2cm}
\includegraphics[width=4cm]{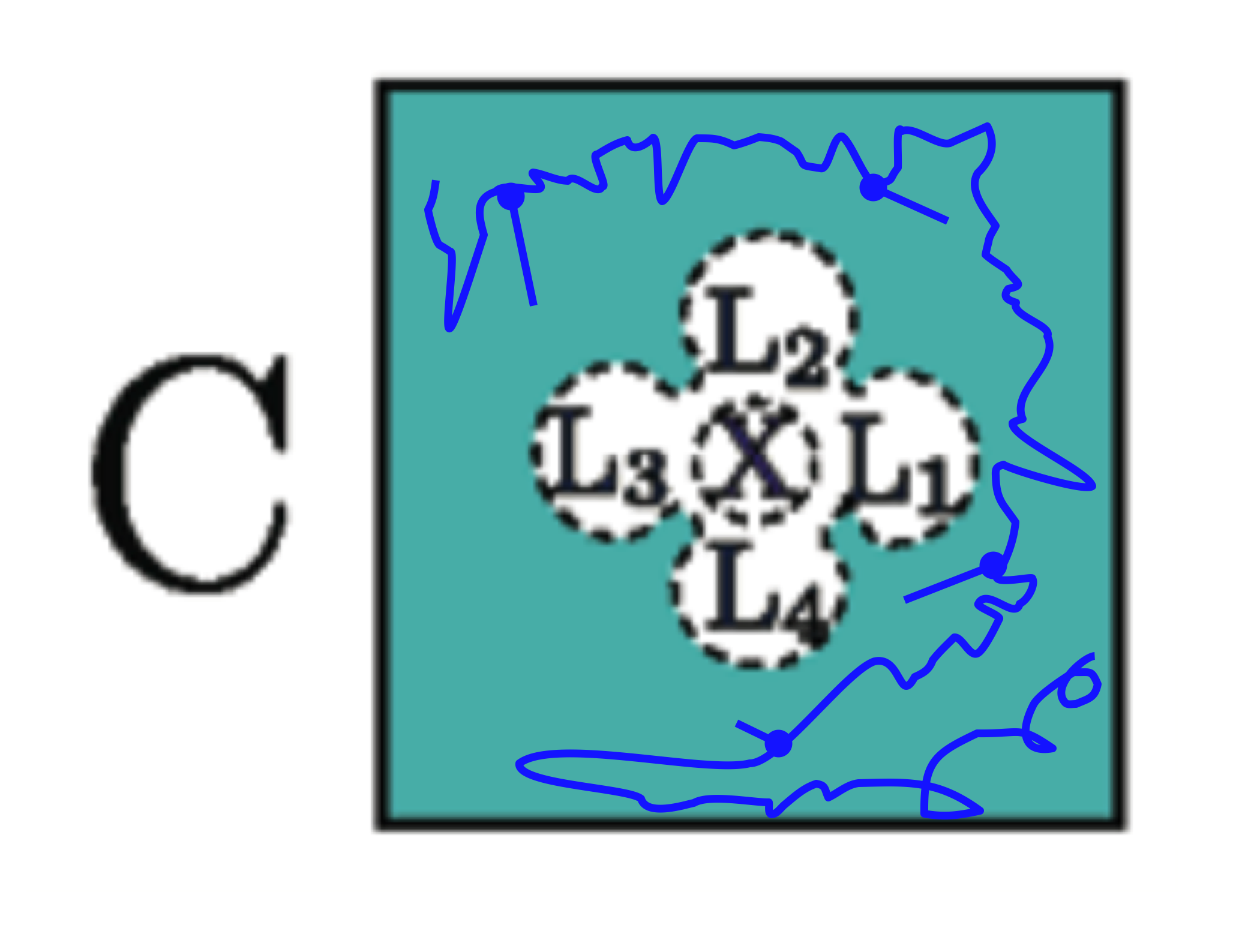}
\label{fig:2c}
\vspace{-0.4cm}
\end{center}
\end{wrapfigure}
State C has both the ion X and the ligands constrained within coordination and excluded volume conditions (surrounded by white shaded area).
The ion and the ligands (surrounded by dotted lines) are fully decoupled from the surroundings (they are not interacting with anything), 
The configurational integral $Z_C$ is,
\begin{equation}
Z_C =  \int d\bX \; d\bY \, I(\bR; C_n) ~ e^{- \beta \left [ \sum_{i=1}^n u^i_{\rm l}(\bx_i) + u_{\rm rr}(\bY) \right ] }  
\end{equation}

\begin{wrapfigure}[4]{r}{4cm}
\begin{center}
\vspace{-1.0cm}
\includegraphics[width=4cm]{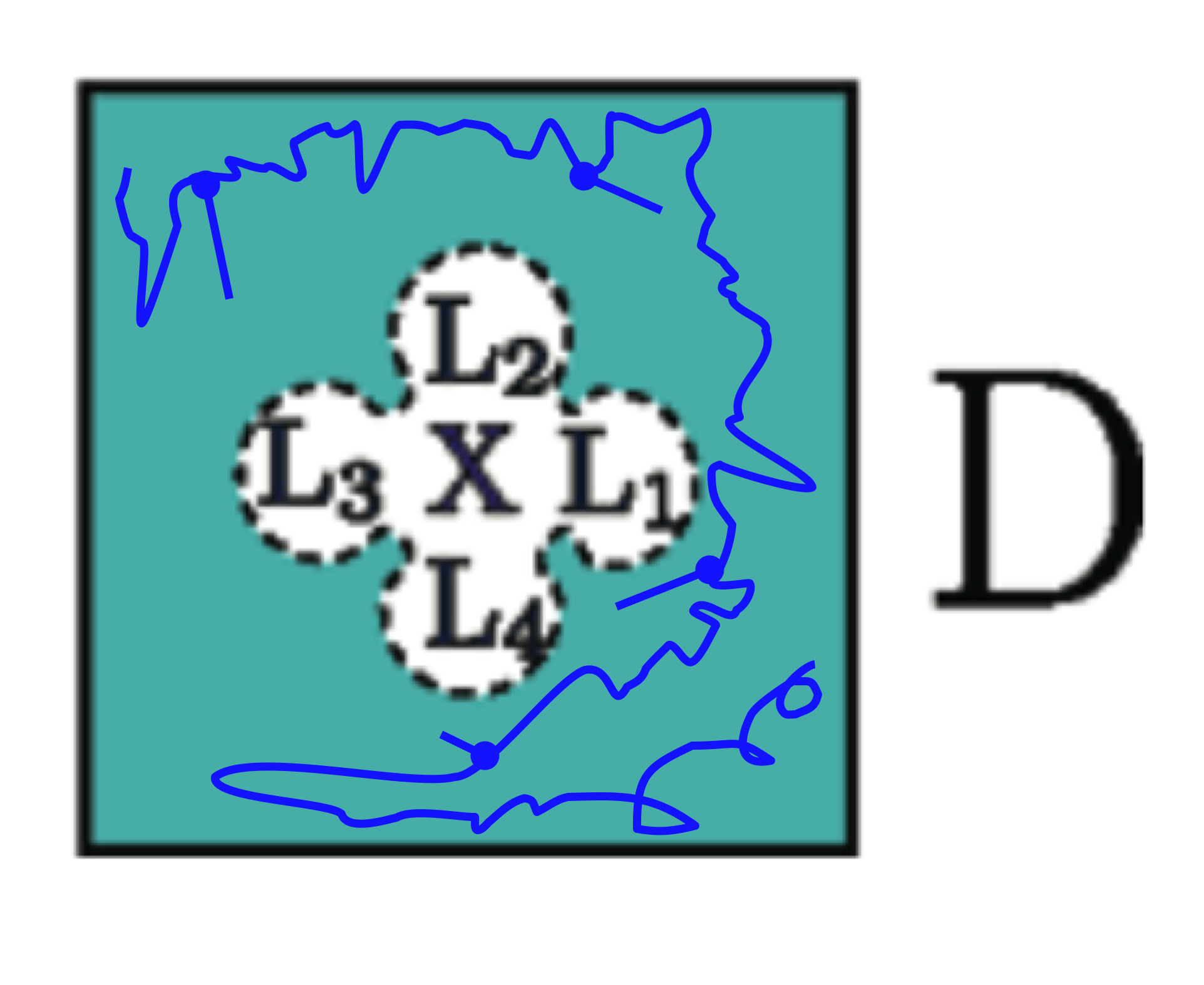}
\label{fig:2d}
\vspace{-0.4cm}
\end{center}
\end{wrapfigure}
State D has both the ion X and the ligands constrained within coordination and excluded volume conditions (surrounded by white shaded area).
They are all surrounded by a single dotted line, 
which implies that the ion and the ligands interact together, but are decoupled from the surrounding.
The configurational integral $Z_D$ is,
\begin{equation}
Z_D =  \int d\bX \; d\bY  \, I(\bR; C_n) ~ e^{- \beta \left [ u_{\rm il}(\bX) + \sum_{i=1}^n u^i_{\rm l}(\bx_i) + u_{\rm ll}(\bX) + u_{\rm rr}(\bY) \right ] }  
\end{equation}

\begin{wrapfigure}[5]{r}{4cm}
\begin{center}
\vspace{-1.0cm}
\includegraphics[width=4cm]{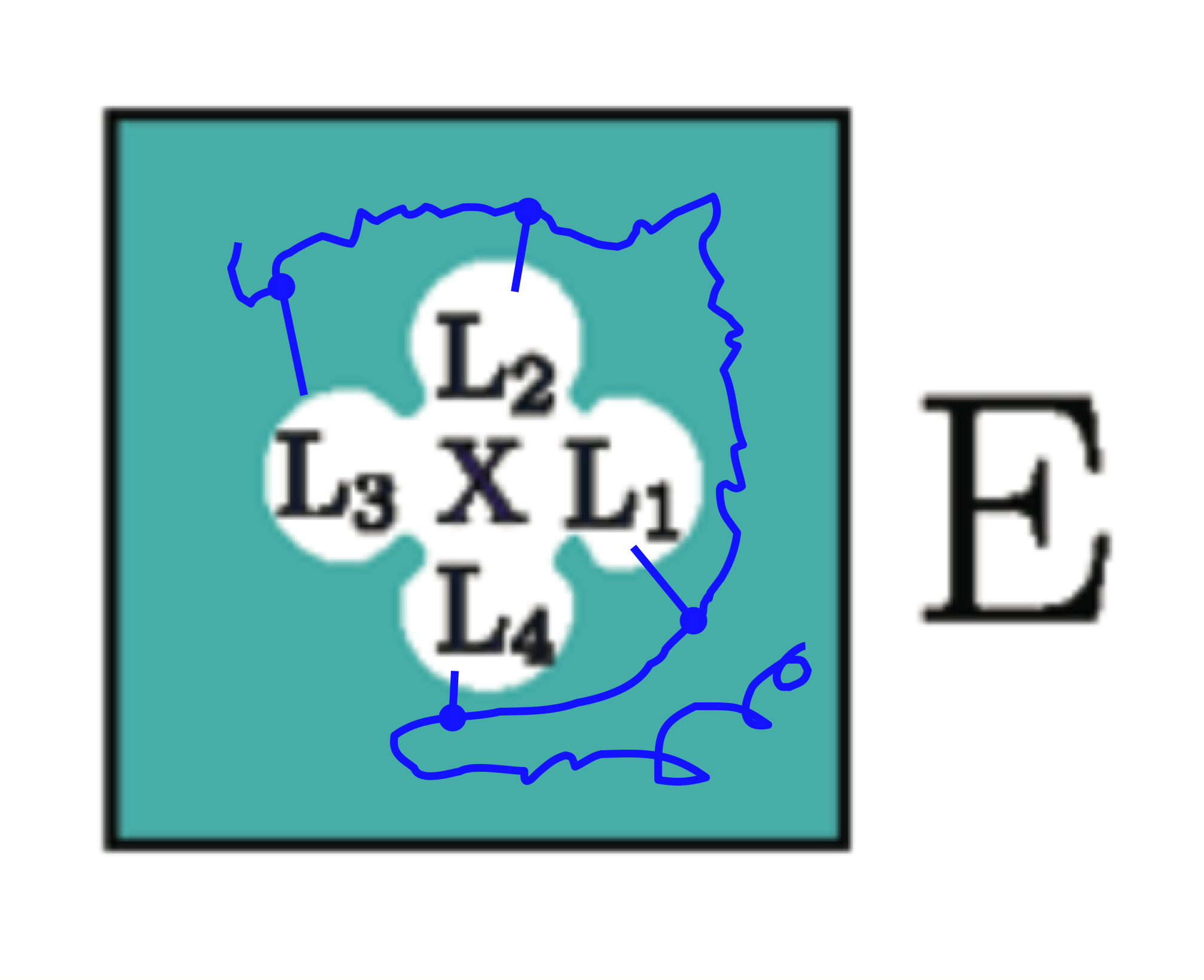}
\label{fig:2e}
\vspace{0.4cm}
\end{center}
\end{wrapfigure}
State E has both the ion X and the ligands constrained within coordination and excluded volume conditions (surrounded by white shaded area).
The ion and the ligands are not surrounded by a dotted line, which implies that they are fully interacting and coupled to the surroundings. 
The configurational integral $Z_E$ is,
\begin{equation}
Z_E =  \int d\bX \; d\bY \, I(\bR; C_n) ~ 
e^{- \beta \left [ u_{\rm il}(\bX) + u_{\rm ir}(\bX,\bY) +  \sum_{i=1}^n u^i_{\rm l}(\bx_i) + u_{\rm ll}(\bX) + u_{\rm lr}(\bX,\bY) +u_{\rm rr}(\bY) \right ] }  
\end{equation}

The meaning of the intermediate steps can be summarized in the following.
At the heart of the cycle, step C$\rightarrow$D is the gas-phase ion-ligands cluster formation that is typical of QCT.
Since this step is an isolated cluster, the plan is to evaluate the configurational
integrals using the rigid rotor, harmonic oscillator (RRHO) approximation, which is often employed in QM 
thermodynamic calculations,
\begin{eqnarray}
{ Z_D \over Z_C } &=& 
{ 
\int d\bX \, I(\bX; C_n) ~ e^{- \beta \left [ u_{\rm il}(\bX) + \sum_{i=1}^n u^i_{\rm l}(\bx_i) + u_{\rm ll}(\bX) \right ] }  
\over 
\int d\bX  \, I(\bX; C_n) ~ e^{- \beta \left [ \sum_{i=1}^n u^i_{\rm l}(\bx_i) \right ] } 
} \nonumber \\
& = & e^{- \beta E_i^{\rm min}(n)} ~ K^{\rm trv}_{\rm i}(n)
\end{eqnarray}
where $E_i^{\rm min}(n)$ is the binding energy for the formation of the isolated cluster (optimal energy-minimized geometry in vacuum), 
While the geometry of the cluster is optimized, it remains 
restricted to a specific $n$ coordination state $C_n$ via the indicator function $I(\bX; C_n)$.
The quantity $K^{\rm trv}_{\rm i}(n)$ corresponds to modified gas phase equilibrium 
translation-rotation-vibration partition functions
for the cluster assembly of one ion with $n$ ligands while they 
are spatially restricted via the function $I(\bX,C_n)$;
note that $I(\bX; C_n)$ was written as a function of $\bX$ without its dependency on the 
coordinates of the rest of the system, $\bY$, since the ion-ligands cluster is assumed to be isolated in the gas phase.

The two steps A$\rightarrow$B at the beginning of the cycle, and E$\rightarrow$F at the end of the cycle, 
consist in imposing the coordination structure $C_n$ to the ion-ligands subsystem.
Step A$\rightarrow$B is a process imposing the coordination structure constraint on the
fully interacting ligands in the presence of a noninteracting ion,
\begin{eqnarray}
{ Z_B \over Z_A } &=&  
{ 
\int d\bX \; d\bY  \, I(\bR; C_n) ~ e^{- \beta [ \sum_{i=1}^n u^i_{\rm l}(\bx_i) + u_{\rm ll}(\bX) + u_{\rm lr}(\bX,\bY) +u_{\rm rr}(\bY)] }  
\over
\int d\bX \; d\bY  \, e^{- \beta [ \sum_{i=1}^n u^i_{\rm l}(\bx_i) + u_{\rm ll}(\bX) + u_{\rm lr}(\bX,\bY) +u_{\rm rr}(\bY)] }  
} \nonumber \\
&=& \langle I(\bX; C_n) \rangle \nonumber \\
&=& P^{\rm apo}(C_n)
\end{eqnarray}
where $P^{\rm apo}(C_n)$ is the probability of finding the ligands coordination structure $C_n$ when the ion is noninteracting.
Step E$\rightarrow$F is a similar process, imposing the coordination structure constraint on the 
fully interacting ion and ligands, 
\begin{eqnarray}
{ Z_F \over Z_E } &=& 
{
\int d\bX \; d\bY ~  
e^{- \beta \left [ u_{\rm il}(\bX) + u_{\rm ir}(\bX,\bY) + \sum_{i=1}^n u^i_{\rm l}(\bx_i) + u_{\rm ll}(\bX) +  u_{\rm lr}(\bX,\bY) +u_{\rm rr}(\bY) \right ] }
\over
\int d\bX \; d\bY  \, I(\bR; C_n) ~ 
e^{- \beta \left [ u_{\rm il}(\bX) + u_{\rm ir}(\bX,\bY) + \sum_{i=1}^n u^i_{\rm l}(\bx_i) + u_{\rm ll}(\bX) +  u_{\rm lr}(\bX,\bY) +u_{\rm rr}(\bY) \right ] } }  \nonumber \\
&=& { 1 \over \langle  I(\bR; C_n) \rangle } \nonumber \\
&=& { 1 \over P^{\rm holo}(C_n) } 
\end{eqnarray}
where $P^{\rm holo}(C_n)$ is the probability of finding the fully interacting ion and ligands in the coordination structure $C_n$.

The two remaining steps, B$\rightarrow$C and D$\rightarrow$E, 
correspond to the process of coupling the subsystem (constrained 
according to the coordination structure $C_n$) to the surrounding.
Step B$\rightarrow$C is the process of de-coupling the ligands constrained 
according to the coordination structure $C_n$ from its surrounding (the ion is noninteracting).
Step D$\rightarrow$E is the process re-coupling the fully interacting ion-ligands complex to its surrounding.
Let us assume--in analogy with the quantity $\Delta W_i$ in Eq.~(\ref{eq:pQCT})--that 
coupling the subsystem (apo or holo) with the surroundings in
steps B$\rightarrow$C and D$\rightarrow$E will yield the free energy contributions 
$\Delta W_{\rm r}^{\rm apo}$ and $\Delta W_{\rm r}^{\rm holo}$, 
\begin{eqnarray}
{ Z_E \over Z_D } &=& 
{
\int d\bX \; d\bY ~  I(\bR; C_n) ~
e^{- \beta \left [ u_{\rm il}(\bX) + u_{\rm ir}(\bX,\bY) +  \sum_{i=1}^n u^i_{\rm l}(\bx_i) + u_{\rm ll}(\bX) + u_{\rm lr}(\bX,\bY) +u_{\rm rr}(\bY) \right ] }
\over
\int d\bX \; d\bY  \, I(\bR; C_n) ~ 
e^{- \beta \left [ u_{\rm il}(\bX) + \sum_{i=1}^n u^i_{\rm l}(\bx_i) + u_{\rm ll}(\bX) +  u_{\rm rr}(\bY) \right ] } }  \nonumber \\
&=& e^{-\beta \Delta W_{\rm r}^{\rm holo} } 
\label{eq:wrong1}
\end{eqnarray}
and
\begin{eqnarray}
{ Z_C \over Z_B } &=& 
{
\int d\bX \; d\bY ~  I(\bR; C_n) ~
e^{- \beta [  \sum_{i=1}^n u^i_{\rm l}(\bx_i) + u_{\rm ll}(\bX)  +u_{\rm rr}(\bY)] }
\over
\int d\bX \; d\bY  \, I(\bR; C_n) ~ 
e^{- \beta [ \sum_{i=1}^n u^i_{\rm l}(\bx_i) + u_{\rm ll}(\bX) + u_{\rm lr}(\bX,\bY) +  u_{\rm rr}(\bY)] } }  \nonumber \\
&=& e^{+\beta \Delta W_{\rm r}^{\rm apo} } 
\label{eq:wrong2}
\end{eqnarray}
(However, see the problems below with the definition of $\Delta W_{\rm r}^{\rm apo}$ and $\Delta W_{\rm r}^{\rm holo}$.)

Using all the contributions from the thermodynamic cycle, 
we can re-express the equilibrium binding constant $K_{\rm b}$ as,
\begin{eqnarray}
K_{\rm b} &=& V  \left ( { Z_F \over Z_A } \right ) \;  e^{\beta G_{\rm bulk}} \nonumber \\
          &=& V 
              \left ( { Z_F \over Z_E } \right ) 
              \left ( { Z_E \over Z_D } \right ) 
              \left ( { Z_D \over Z_C }\right ) 
              \left ( { Z_C \over Z_B }\right ) 
              \left ( { Z_B \over Z_A }\right ) \; 
e^{\beta G_{\rm bulk}} \nonumber \\
&=&  \left \{ V e^{- \beta E_i^{\rm min}(n)} ~ K^{\rm trv}_{\rm i}(n) ~ e^{\beta G_{\rm bulk}} \right \} ~ \times 
     \left \{ { P^{\rm apo}(C_n)  \over P^{\rm holo}(C_n) } ~  
              e^{-\beta [ \Delta W_{\rm r}^{\rm holo} - \Delta W_{\rm r}^{\rm apo} ] }  \right \}
\end{eqnarray}
The first curly bracket is only concerned with the gas-phase QCT cluster binding free energy $K^{\rm trc}_{\rm i}(n)$ 
and the hydration free energy of the ion $G_{\rm bulk}$.
In the treatment of a binding site via the QCT for inhomogeneous systems, 
this part is clearly meant to be a central quantity that must be
evaluated by high-level quantum mechanical methods with energy minimization and a RRHO approximation.
However, while the contribution from the ion-ligands cluster is important, it is critical to include 
the other terms in order to account for the properties of a real protein binding site.

All external effects (e.g., protein conformational preferences, bulk system composition, applied membrane electric field, 
surface tension, pressure, etc) are supposed to be taken into account by the terms contained within the second curly brackets.
Here, these effects are expressed in the form of a probability for formation of each state
must be included in the probabilities $P^{\rm apo}(C_n)$ and $P^{\rm holo}(C_n)$, 
and in the free energy contributions  $\Delta G_{\rm r}^{\rm apo}$ and $\Delta G_{\rm r}^{\rm holo}$.
In principle, the probabilities $P^{\rm apo}(C_n)$ and $P^{\rm holo}(C_n)$ 
can be calculated by using simulations of a realistic all-atom model of the entire system, though
it is worth stressing that these probabilities cannot be evaluated by considering only an isolated ion-ligands cluster in the gas phase.
In contrast, the free energy contributions  $\Delta W_{\rm r}^{\rm apo}$ and $\Delta W_{\rm r}^{\rm holo}$,
which correspond to the coupling of the subsystem to its surroundings, are extremely problematical. 
These quantities are ``defined'' by Eqs.~(\ref{eq:wrong1}) and (\ref{eq:wrong2}),
following the schematic notation with ``dotted line and white shading'' adopted in the Fig. 2 of ref \cite{Rogers-2011}.
In the case of ion solvation in a bulk liquid, which is the normal route in the development of QCT \cite{Roux-QCT-2010}, 
such a decoupling step is not problematic because the
surroundings is constituted by solvent molecules that are free to adapt to the ion-ligands cluster.
Hence, the quantity $\Delta W_i$ in Eq.~(\ref{eq:pQCT}) is perfectly well defined for a cluster immersed into a bulk liquid.
However, in the case of ligands that are covalently attached to a protein structure, this thermodynamic route is impractical.
In reality, a protein structure is expected to be extensively disrupted when the ligand-rest coupling 
$u_{\rm lr}(\bX,\bY)$ is switched off (indicated schematically by a wavy blue line in the schematic representations of state C and D).
Therefore, even though the ion and the ligands may remain spatially
restricted by the function $I(\bR; C_n)$, the rest of the protein will distort and
probably unfold as the subsystem of ligands is decoupled from the rest of the structure.
Expressions such as Eqs.~(\ref{eq:wrong1}) and (\ref{eq:wrong2}) can be formally written, 
but they are essentially meaningless an unusable in practice.
In conclusion, the free energy contributions defined via the steps B$\rightarrow$C and D$\rightarrow$E,
cannot be used to carry out a practical calculation based on a realistic atomic model 
for the purpose of evaluating the effect of the protein structure on an ion-ligands subsystem.

\subsection*{c) Small System Grand Canonical Ensemble (SSGCE)}

Bostick and Brooks presented a statistical mechanical framework for deconstructing the determinants of 
selective ionic complexation in protein binding sites \cite{Bostick-2009}.
The theoretical developments lean on the concept of SSGCE considered by Reiss and co-workers \cite{Reiss-1998},
and start by considering the statistical distribution function of the ligands around an 
ion interacting with a polypeptide comprising $N$ possible coordinating ligands. 
The resulting framework is closely related to QCT for an ion solvated in a bulk liquid \cite{Roux-QCT-2010}.
However, a careful examination reveals fundamental problems in the case of an ion bound to a protein binding site. 

The notation of ref \cite{Bostick-2009} is kept for the sake of clarity.
The $N$ ligands, located at position $\{\br_1,\dots,\br_N\}$, are assumed to be chemically identical (e.g., they are all backbone carbonyl oxygens). 
The potential energy of the system is $U(\br_1,\dots,\br_N, \bR)$, where $\bR$ denotes additional degrees of freedom from the solvent.
Without loss of generality it is assumed that the ion is  fixed at the origin 
(in the following, the fixed ion is implicitly included in the configurational integrals but is omitted in the notation for clarity).  
An inner region is defined as a spherical subvolume $v$ within a distance $r_{\rm c}$ away from the ion, 
and a counting function, $c(\br_1,\dots,\br_N)$, is introduced to monitor the number of ligands within the inner region,
$c (\br_1,\dots,\br_N) = \sum_{i=1}^N ~ H(|\br_i|-r_{\rm c})$ (for any configuration, the $c$ is equal to the number of ligands in the inner subvolume).
The spherical subvolume is analogous to the inner shell that is defined in pQCT (see Fig. 1A, left).
Defining the discrete Kroenecker delta function $\delta_{c,n}$, 
the probability to have exactly $n$ ligands in the inner region, $P^{CR}(n)= \langle \delta_{c(\br_1,\dots,\br_N), n} \rangle$, 
is written as,
\begin{equation}
P^{CR}(n) = { 1 \over Z } \int d\br_1 \dots \int d\br_N \int d\bR ~ \delta_{c(\br_1,\dots,\br_N), n} ~ e^{-\beta U(\br_1,\dots,\br_N, \bR) }    
\label{eq:A4} 
\end{equation} 
where $Z$ is the unrestricted configurational integral,
\begin{equation}
Z =    \int d\br_1 \dots \int d\br_N \int d\bR  ~  e^{-\beta U(\br_1,\dots,\br_N, \bR) } 
\end{equation}
Eq.\,(\ref{eq:A4}) can be expanded to explicitly display  the contribution from all  possible permutations of 
the $N$ ligands such that exactly $n$ of those are located within the inner region, 
\begin{eqnarray}          
P^{CR}(n) &=& { 1 \over Z } \int_{\rm in} d\br_1 \dots \int_{\rm in} d\br_n \int_{\rm out} d\br_{n+1} \dots \int_{\rm out} d\br_N \int d\bR ~  e^{-\beta U(\br_1,\dots,\br_N, \bR) }  \nonumber \\
          & & \hspace{-0.8in} +  { 1 \over Z }\int_{\rm in} d\br_1  \dots \int_{\rm in} d\br_{n-1} \int_{\rm out} d\br_n \int_{\rm in} d\br_{n+1} \int_{\rm out} d\br_{n+2} \dots \int_{\rm out} d\br_N ~  e^{-\beta U(\br_1,\dots,\br_N, \bR) }  \nonumber \\
           & & \hspace{-0.8in} +   { 1 \over Z }\int_{\rm in} d\br_1  \dots \int_{\rm in} d\br_{n-1} \int_{\rm out} d\br_n \int_{\rm out} d\br_{n+1} \int_{\rm in} d\br_{n+2} \int_{\rm out} d\br_{n+3} \dots \int_{\rm out} d\br_N \int d\bR ~  e^{-\beta U(\br_1,\dots,\br_N, \bR) }  \nonumber \\
          & & \hspace{-0.8in}+ \dots   
\label{eq:permutations}
\end{eqnarray}
There are $N!/n!(N-n)!$ permutations of the $N$ ligands such that exactly $n$ are located inside the inner region.   
If the ligands were solvent molecules in a liquid, then all those permutations would yield equivalent expressions of equal weight.
One could choose one of the possible permutation, 
for example the first term in Eq.\,(\ref{eq:permutations})  with ligands $1$ to $n$ within the inner region and ligands $n+1$ to $N$ outside, 
and then write  the probability to have exactly $n$ ligands in the inner region as 
\cite{Reiss-1998, QCT-Pratt-1998, Woo-2004, Deng-2008},
\begin{equation}          
P^{CR}(n) =  { N! \over n!(N-n)! } { 1 \over Z } \int_{\rm in} d\br_1 \dots \int_{\rm in} d\br_n \int_{\rm out} d\br_{n+1} \dots \int_{\rm out} d\br_N \int d\bR ~  e^{-\beta U(\br_1,\dots,\br_N, \bR) }   
\label{eq:A7}
\end{equation}
While the relation between SSGCE and QCT were not noted by the authors, the mathematical steps corresponding to 
Eqs.~(\ref{eq:A4}-\ref{eq:A7}) are identical to those used to derive QCT for an ion solvated in a bulk liquid 
(see Eqs. 6-13 in ref \cite{Roux-QCT-2010}).

\begin{figure}[t!]
\begin{center}
\includegraphics[width=6.5in]{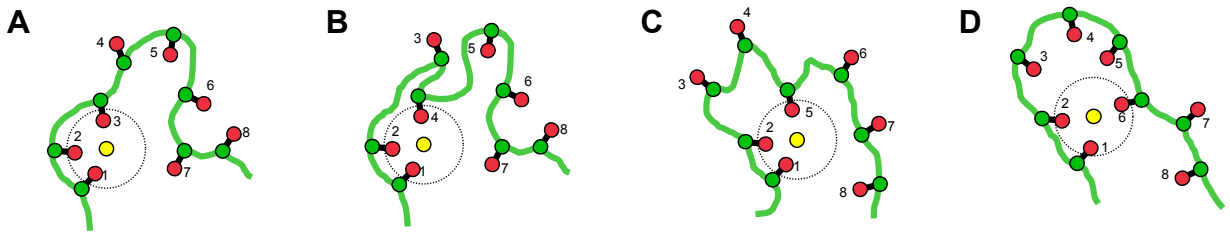}
\end{center}
\caption{   
\footnotesize
Schematic illustration of configurations of an ion coordinated by $n$ ligands bonded to a polypeptide chain 
of $N$ ligands with a fixed topological structure in the Small System Grand Canonical Ensemble (SSGCE) formulation.
The ion is depicted as a yellow circle placed at the origin, 
the ligands are depicted as green-red dumbbells, 
the covalent topological structure of the polypeptide backbone with its predetermined connectivity is depicted schematically as a thick green line, 
and the solvent molecules are not shown.
The inner  subvolume region is shown as a thin dotted line.
For the sake of the figure, $N=8$ and $n=3$. 
In A is shown a configuration with ligands 1-3 located within the inner subvolume, and ligands 4-8 located outside the inner subvolume region. 
The energy of the configuration is $U(\br_1,\br_2, \br_3, \br_4, \br_5, \br_6, \br_7, \br_8)$.
In B is shown a configuration obtained from A by swapping the coordinates of ligands 3 and 4; ligands 1, 2 and 4 are within the inner subvolume region, and ligands 3, 5, 6, 7, and 8 are located outside the inner subvolume region.
The energy of the configuration is $U(\br_1,\br_2, \br_4, \br_3, \br_5, \br_6, \br_7, \br_8)$, different from the energy of the configuration in A.
Configurations A and B would strictly be equivalent if the polypeptide backbone was removed and the ligands were separate molecules.
Configurations A-D, all with $n=3$ ligands within the subvolume, provide additional examples of contributions to  the sum over permutations expressed in Eq.\,(\ref{eq:permutations}); 
A-C represents typical contribution to the first, second and third configurational integral in Eq.\,(\ref{eq:permutations}).  
All four configurational integral pictured schematically in A-D is expected to have a different statistical weight.
Because configurational integral associated with A-D are expected to have different statistical weights, 
it is not possible in general to represent the total sum over all possible permutations of Eq.\,(\ref{eq:permutations}) in terms of a combinatorial factor as in Eq.\,(\ref{eq:A7}).
}
\end{figure}

The critical issue that is of concern is the use of Eq.\,(\ref{eq:A7}) to represent the probability 
of finding $n$ protein ligands within the inner shell region of an ion (this is Eq. (A7) in ref \cite{Bostick-2009}).
As illustrated schematically in Fig. 2, the use of the combinatorial factor $N!/n!(N-n)!$
in Eq.\,(\ref{eq:A7}) to represent the total sum over all possible permutations expressed in 
Eq.\,(\ref{eq:permutations}) is strictly valid only if the Boltzmann factor is {\em invariant} upon the exchange (swapping) of ligand coordinates.
In a liquid, permutation of any pair of chemically identical molecules $i$ and $j$ yields the same potential energy, 
$U(\dots,\br_i,\dots,\br_j,\dots) =   U(\dots,\br_j,\dots,\br_i,\dots)$, 
and it is possible to write Eq.\,(\ref{eq:A7}) with the combinatorial factor \cite{Reiss-1998, QCT-Pratt-1998, Woo-2004, Deng-2008}.
But such a simplification is invalid when the ligands are covalently linked to a polypeptide.
Even if ligands $i$ and $j$ are ``chemically identical'', swapping their coordinates should not generate an equivalent 
configuration with the same potential energy and Boltzmann weight because they are covalently attached to the polypeptide chain.
The lack of invariance upon exchange of identical ligands is automatically satisfied, by construction, 
if the energy of the polypeptide is represented on the basis of a classical molecular mechanical force field 
with a fixed list of chemical bonds \cite{CHARMM-2009}.  
But more generally, even if the energy of the polypeptide were represented on the basis of
a quantum mechanical Born-Oppenheimer energy surface that is formally invariant upon exchange of identical nuclei, 
restricting the configurational integral of Eq.~(\ref{eq:A7}) to a single state of covalent topological connectivity
would still be necessary to formulate a meaningful statistical mechanical theory of ion binding to a protein site.
Doing otherwise, i.e., constructing $P^{CR}(n)$ from a superposition of all configurational states including 
those obtained by swapping the coordinates of identical nuclei, implies that one is averaging over processes that 
are actually slower than any macroscopic observation timescale.
Bond breaking and forming can occur in a polymer, but under normal conditions at room temperature, atoms and residues along a polypeptide chain do 
not spontaneously swap their position (except labile hydrogens). 
For example,  it is possible to isotopically label the atoms of any carbonyl group in the gramicidin channel
to perform NMR experiments, and such labeling can remain stable for months \cite{Cross-1993b}.  
The permutation of identical ligands implicit in Eq.\,(\ref{eq:A7}) is incorrect because a
valid statistical mechanical treatment should only sum over the microstates that are
accessible within a reasonable observation time (see pp. 42-43 in ref \cite{Hill-1962}).

The statistical mechanical framework presented by Bostick and Brooks was designed to show how the coordination structure of 
an ion solvated by a liquid provides the key elements needed to understand the selectivity displayed by a protein binding site.
To firm up this view, the authors analyzed MD simulations of simplified model systems comprising an ion immersed in a generic fluid
of unrestrained ligands in the context of their statistical mechanical framework. 
The fluid of unrestrained ligands, referred to as the ``soup of ligands'' or ``hypothetical fluid of coordinators''(HFC), serves as a key reference in the analysis.
But the significance of the analysis based on the HFC is unclear.
The HFC, rooted in Eq. (A7), assimilates the combinatorial factor associated with the configurations 
of an ion solvated in a liquid to that of an ion coordinated by the covalently linked ligands in a protein binding site.

By incorrectly treating protein ligands as identical solvent molecules that are free to exchange with one another 
in the bulk phase, the theoretical development presented in ref \cite{Bostick-2009} 
creates the impression that information directly relevant to an ion binding site in a protein can be extracted from the properties of the HFC.
In reality, the link between an ion-selective protein binding and the HFC is unclear.

\section*{3. Summary}

While the concept of a reduced model is intuitively obvious, many choices are possible, 
leading to important differences in formulations and implementations.
The general idea of a reduced model is that the ``near'' degrees of freedom must be treated explicitly
while the influence of the surrounding must be implicitly incorporated.
Frameworks based on quasi-chemical theory (QCT) \cite{Varma-2007,Varma-2008b,Rogers-2011}
and on the small system grand canonical ensemble (SSGCE) development \cite{Bostick-2009}
have previously been proposed to achieve such a reduction of ion binding sites in macromolecules into reduced models.
However, a number of problems were discovered with these frameworks.
Ultimately, the major shortcomings of approaches based on QCT or SSGCE as formulated until now lie in 
their {\em inability to account for the influence of the covalently linked protein structure}. 
One consequence is that these formulations fail to draw a clear distinction between 
ion-selective protein sites that are either very rigid or very flexible \cite{Yu-PNAS-2010}.
Developing a valid statistical mechanical theory of reduced models for ion binding sites in 
macromolecules remains an important and valuable goal for future efforts.

\subsection*{Acknowledgments}

The work was funded by grant GM-62342 from the National Institute of Health (NIH).  
Helpful discussion with Chris. N. Rowley are gratefully acknowledged.

{\footnotesize
\bibliography{/Users/roux/papers/biblio/thesis}
\bibliographystyle{pnas-bolker}
}
\end{document}